\documentclass[final,5p,times]{amsci}

\usepackage{amssymb,amsfonts,amsmath,amstext,amsgen,amsopn,amsxtra,indentfirst,lscape}

\journal{Volume 102, Number 3, Pages 174--177 (May/June 2014 Issue)}

\begin{document}

\begin{frontmatter}

\title{The Nature of Scientific Proof in the Age of Simulations}

\author{Kevin Heng}
\ead[csh]{kevin.heng@csh.unibe.ch}
\address{University of Bern \\ Center for Space and Habitability \\ Sidlerstrasse 5, CH-3012, Bern, Switzerland}

\begin{abstract}
\textit{Is numerical mimicry a third way of establishing truth?\vspace{0.05in}\\
{\scriptsize Kevin Heng received his M.S. and Ph.D. in astrophysics from the Joint Institute for Laboratory Astrophysics (JILA) and the University of Colorado at Boulder. He joined the Institute for Advanced Study in Princeton from 2007 to 2010, first as a Member and later as the Frank \& Peggy Taplin Member. From 2010 to 2012 he was a Zwicky Prize Fellow at ETH Z\"{u}rich (the Swiss Federal Institute of Technology). In 2013, he joined the Center for Space and Habitability (CSH) at the University of Bern, Switzerland, as a tenure-track assistant professor, where he leads the Exoplanets and Exoclimes Group. He has worked on, and maintains, a broad range of interests in astrophysics: shocks, extrasolar asteroid belts, planet formation, fluid dynamics, brown dwarfs and exoplanets. He coordinates the Exoclimes Simulation Platform (ESP), an open-source set of theoretical tools designed for studying the basic physics and chemistry of exoplanetary atmospheres and climates (www.exoclime.org). He is involved in the CHEOPS (Characterizing Exoplanet Satellite) space telescope, a mission approved by the European Space Agency (ESA) and led by Switzerland. He spends a fair amount of time humbly learning the lessons gleaned from studying the Earth and Solar System planets, as related to him by atmospheric, climate and planetary scientists. He received a Sigma Xi Grant-in-Aid of Research in 2006.} \vspace{0.05in}\\
Edited by Katie Burke.  \texttt{www.americanscientist.org }
} 
\end{abstract}

\end{frontmatter}

Empiricism lies at the heart of the scientific method.  It seeks to understand the world through experiment and experience.  This cycle of formulating and testing falsifiable hypotheses has amalgamated with a modern form of rationalism---the use of reasoning, mathematics and logic to understand nature.  These schools of thought are couched in centuries of history and, until recently, remained largely distinct.  Proponents of empiricism include the eighteenth-century Scottish philosopher, David Hume, who believed in a subjective, sensory-based perception of the world.  Rationalism is the (mistaken) belief that the use of reasoning alone is sufficient to understand the natural world, without any recourse to experiment.  Its roots may be traced to the Greek philosophers, Aristotle, Plato and Pythagoras; its more modern proponents include Kant, Leibniz and Descartes. 

A clear example of both practices at work is in the field of astronomy and astrophysics. Astronomers discover, catalogue, and attempt to make sense of the night sky using powerful telescopes. Astrophysicists mull over theoretical ideas, form hypotheses, make predictions for what one expects to observe, and attempt to discover organizing principles unifying astronomical phenomena. Frequently, researchers are practitioners of both subdisciplines. 

Problems in astrophysics---and physics, in general---may often be rendered tractable by concentrating on the characteristic length, time, or velocity scales of interest. When trying to understand water as a fluid, it is useful to treat it as a continuous medium rather than as an enormous collection of molecules, because it makes it vastly easier to visualize (and compute) its macroscopic behavior. Although the Earth is evolving on geological time scales, its global climate is essentially invariant from one day to the next (and hence the difficulty in explaining the urgency of climate change to the public). The planets of the Solar System do not orbit a static Sun, as it performs a ponderous wobble about its center of mass due to their collective gravitational tug, but it is often sufficient to visualize it as being so. The Milankovitch cycles cause the eccentricity and obliquity of the EarthÕs orbit to evolve over hundreds of thousands of years, but they are essentially constant over a human lifetime. This separation of scales strips a problem down to its bare essence, allowing one to gain insight into the salient physics at the scale of interest.  

Multiscale problems, on the other hand, do not lend themselves to such simplification. Small disturbances in a system might show up as big effects across a myriad of sizes and time scales. Structures on very large scales ``talk" to features on very small scales and vice versa. For example, a grand challenge in astrophysics is planet formation---to predict the diversity of exoplanets forming around a star, starting from a primordial cloud of gas and dust. Planet formation is an inherently multiscale problem: Uncertainties on microscopic scales, such as how turbulence and the seed particles of dust grains are created, hinder our ability to predict the outcome on celestial scales. Many ``real-life" problems in biology, chemistry, physics, and the atmospheric and climate sciences are multiscale. 

By necessity, a third, modern way of testing and establishing scientific truth---in addition to theory and experiment---is via simulations, the use of (often large) computers to mimic nature. It is a synthetic universe in a computer. One states an equation (or several) describing the physical system being studied, programs it into a computer, and marches the system forward in space and time. If all of the relevant physical laws are faithfully captured, then one ends up with an emulation---a perfect, \textit{The Matrix}-like replication of the physical world in virtual reality. 

In astronomy and astrophysics, this third way has come into its own, largely due to the unique status of astronomy as an experimental science. Unlike other, laboratory-based disciplines, astronomers may not exert full control over their experiments---one simply cannot rearrange objects in the sky. Astronomical phenomena often encode information about a subpopulation of a class of objects at a very specific moment in their evolution. To understand the entire population of a class of objects across cosmic time requires large computer simulations of their formation and evolution. Examples of different classes of objects include exoplanets, stars, black holes, galaxies and even clusters of galaxies. The hope is that these simulations lead to ``big picture" understanding that unifies seemingly unrelated astronomical phenomena. 

During the 1940s through the 1980s, the late, distinguished Princeton astrophysicist Martin Schwarzschild was one of the first to use simulations to gain insights into astronomy, harnessing them to understand the evolution of stars and galaxies. Schwarzschild realized that the physical processes governing stellar structure are nonlinear and not amenable to analytical (``pencil and paper") solution since it requires an understanding of the physics of nuclear burning, while galaxies are hardly perfect spheres, and proceeded to investigate them using numerical solutions generated by large computers (at that time). 

Both lines of inquiry have since blossomed into respected and full-fledged subdisciplines in astrophysics. Nowadays, an astrophysicist is as likely to be found puzzling over the engineering of complex computer code as he or she is to be found fiddling with mathematical equations on paper (or chalkboard). 

From the 1990s to the present, the approach of using computer simulations for testing hypotheses flourished. As technology advanced, astronomical datasets became richer, motivating the need for more detailed theoretical predictions and interpretations. Computers became more prevalent and faster, alongside rapid advances in the algorithmic techniques developed by computational science. Inexorably, the calculations produced by large simulations evolved to resemble experimental datasets in size, detail and complexity. Computational astrophysicists now come in three variants: engineers to build the code, researchers to formulate hypotheses and design numerical experiments, and yet others to process and interpret the resulting massive output. Supercomputing centers function almost like astronomical observatories. For better or worse, this third way of establishing scientific truth appears to be here to stay.

In a series of lunchtime conversations with astrophysicist Piet Hut of the Institute for Advanced Study in Princeton, I discovered that we were both concerned about the implications of these ever-expanding simulations. Computational astrophysics has adopted some of the terminology and jargon traditionally associated with the experimental sciences. Simulations may legitimately be regarded as numerical experiments, along with the assumptions, caveats, and limitations associated with any traditional, laboratory-based experiment. Simulated results are often described as being ``empirical", a term usually reserved for natural phenomena rather than numerical mimicries of nature. Simulated data are referred to as ``data sets", seemingly placing them on an equal footing with observed natural phenomena. 

The Millennium Simulation Project, designed and executed by the Max Planck Institute for Astrophysics in Munich, Germany, provides a pioneering example of such an approach. It is a massive simulation of a universe in a box, elucidating the very fabric of the cosmos. The datasets generated by these simulations are so widely used that entire workshops are organized around them. Mimicry has supplanted astronomical data. 

It is not far-fetched to say that all theoretical studies of nature are approximations. There is no single equation that describes all physical phenomena in the universe---and even if we could write one down in principle, solving it would be prohibitive, if not downright impossible. The equations we study as theorists are merely approximations of nature. Schr\"{o}dinger's equation describes the quantum world in the absence of gravity. The Navier-Stokes equation is a macroscopic description of fluids. Newton's equation describes gravity accurately under terrestrial conditions, superceded only by Einstein's equations under less familiar conditions. 

To understand the orbital motion of exoplanets around distant stars, it is mostly sufficient to only consider Newtonian gravity. To understand the appearance of these exoplanetsÕ atmospheres requires approximating them as fluids and understanding the macrosopic manifestations of the quantum mechanical properties of the individual molecules (their absorption and scattering properties). Each of these governing equations is based on a law of Nature---the conservation of mass, energy or momentum (or some other generalized, more abstract quantity such as potential vorticity).  One selects the appropriate governing equation of nature and solves it in the relevant physical regimes, thus creating a model. A model captures a limited set of salient properties of a physical system. The term itself is widely abused---a ``model" that is not based on a law of nature has little right to be called one.

A fundamental limitation of any simulation is that there is a practical limit to how finely one may slice space and time in a computer (the ``resolution"), such that the simulation completes within a reasonable amount of time (say, within the duration of one's Ph.D thesis). For multiscale problems, there will always be phenomena operating on scales smaller than the size of oneÕs simulation pixel. Astrophysicists term these ``subgrid physics"---literally physics happening below the grid of the simulation. This difficulty of simulating phenomena from microscopic to macroscopic scales, across many, many orders of magnitude in size, is known as a ``dynamic range" problem. 

As computers become more powerful, one may always run simulations that explore a greater range of sizes and discretize space and time ever more finely, but in multiscale problems there will always be unresolved subgrid phenomena. Astrophysics and climate science appear to share this nightmare. In simulating the formation of galaxies, the birth, evolution, and death of stars are determining the global appearance of these synthetic galaxies themselves. Galaxies typically span tens of thousands of light years across, whereas stars operate on scales that are roughly a hundred billion times smaller. 

The climate of Earth appears to be significantly influenced by clouds, which both heat and cool the atmosphere. On scales of tens to hundreds of kilometers, it is the imperfect cancellation between these two effects that matters.  To get the details of this cancellation correct, we need to understand how the clouds formed and how their emergent properties developed, which ultimately requires an intimate understanding of how the microscopic seed particles of clouds were first created (``nucleation"). Remarkably, uncertainties about cloud formation on such fine scales are hindering our ability to predict whether a given exoplanet is potentially habitable. Cloud formation remains a largely unsolved puzzle across several scientific disciplines. In both examples, it remains challenging to simulate the entire range of phenomena, both due to the prohibitive amount of computing time needed and our incomplete understanding of the physics involved on smaller scales.  

Another legitimate concern is the use of simulations as ``black boxes" to churn out results and generate seductive graphics or movies without deeply questioning the assumptions involved. For example, simulations involving the Navier-Stokes equation often assume a ``Newtonian fluid"---one that retains no memory of what was done to it in the past and offers more resistance or friction when layers of it are forced to slide past one another. Newtonian fluids are a plausible starting point for a rich variety of simulations, ranging from planetary atmospheres to accretion disks around black holes. Curiously, several common fluids are non-Newtonian. Dough is an example of a fluid with a memory of its past states, while ketchup tends to become less viscous when it is increasingly deformed. Attempting to simulate these fluids using a Newtonian assumption is an exercise in futility. 

To use a simulation as a laboratory, one has to understand how to break it---otherwise, one may mistake an artifact as a result. In approximating continua as being discrete, one has to pay multiple penalties. Spurious oscillations or enhanced viscosity that are artifacts of this procedure may easily be misinterpreted as being physically meaningful.  Simply put, when one slices up space and time in a simulation, it may introduce features that look like real waves or make the fluid more viscous in an artificial way. The conservation of mass, momentum, and energy---cornerstones of theoretical physics---may no longer be taken for granted in a simulation and depends on the numerical scheme being employed, even if the governing equation conserves all of these quantities perfectly on paper. 

Despite these concerns, a culture of ``bigger, better, faster" is prevailing. It is not uncommon to hear discussions centered on how one can make one's code more complex and run even faster on a mind-boggling number of computing cores. It is almost as if gathering exponentially increasing amounts of information will automatically translate into knowledge, that the simulated system attains self-awareness. As terabytes upon terabytes of information are being churned out by ever more massive simulations, the gulf between information and knowledge is widening. We appear to be missing a set of guiding principles---a ``meta-computational astrophysics", for lack of a better term. 

Questions for metacomputational astrophysics include: Is scientific truth more robustly represented by the simplest or the most complex model? (Many would say simplest, but this view is not universally accepted.) How may we judge when a simulation has successfully approximated reality in some way? (The visual inspection of a simulated image of, say, a galaxy versus one obtained with a telescope is sentimentally satisfying, but objectively unsatisfactory.) When is ``bigger, better, faster" enough? Does one obtain an ever better physical answer by simply ramping up the computational complexity? 

An alternative approach to ``bigger, better, faster" is to construct a ``model hierarchy"---a suite of models of varying complexity that develops understanding in steps, allowing each physical effect to be isolated. Model hierarchies are standard practice in climate science. Focused models of microprocesses (turbulence, cloud formation, etc) buttress global simulations of how the atmosphere, hydrosphere, biosphere, cryosphere and lithosphere interact.

With increasingly complex simulations, there are also questions surrounding the practice of science. It is not unheard of to encounter published papers in astrophysics where insufficient information is provided for the reproduction of simulated results. Frequently, the computer codes used to perform these simulations are proprietary and complex enough that it would take years and the dedicated efforts of a research team to completely recreate them. Scientific truth is monopolized by a few and dictated to the rest. Is it still science if the results are not readily reproducible? (Admittedly, ``readily" has a subjective meaning.) 

There are also research groups or individuals who take the more modern approach of making their codes open source. This has the tremendous advantage that the task of scrutinizing, testing, validating, and debugging the code no longer rests upon the shoulders of an individual, but of the entire community. Some individuals believe that this amounts to giving away trade secrets, but there are notable examples of researchers whose careers have blossomed partly because of influential computer codes they have made freely available. 

A pioneer in this regard is Sverre Aarseth, a Cambridge astrophysicist who wrote and freely gave away codes that computed the evolution of astronomical objects (planets, stars, etc) under the influence of gravity (``N-body" codes). Jim Stone of Princeton and Romain Teyssier of Z\"{u}rich are known for authoring a series of codes that solve the equations of magnetized fluids and they have been used to study a wide variety of problems in astrophysics. Volker Springel of Heidelberg made his mark via the Millenium Simulation Project. In all of these cases, the publicly available computer codes became influential because other researchers incorporated them into their repertoire and they eventually became part of the astrophysical lexicon. 

A related issue is falsifiability. If a physical system is perfectly understood, it comes with no freedom of specifying model inputs. Technically, astrophysicists term these ``free parameters". Quantifying how the sodium atom absorbs light provides a fine example---it is a triumph of quantum physics that such a calculation requires no free parameters. In large-scale simulations, there are always physical aspects that are poorly or incompletely understood and need to be mimicked by approximate models that specify free parameters. Often, these pseudo-models are not based on fundamental laws of physics, but consist of ad hoc functions calibrated on experimental data or smaller-scale simulations, which may not be valid in all physical regimes. 

An example is the planetary boundary layer on Earth, which arises from the friction between the atmospheric flow and the terrestrial surface and is an integral part of the climatic energy budget. The exact thickness of the planetary boundary layer depends on the nature of the surface---whether it is an urban area, grasslands, or ocean matters. Such complexity cannot be directly and feasibly computed in a large-scale climate simulation. Hence, one needs experimentally measured prescriptions for the thickness of this layer as inputs for the simulation. To unabashedly apply these prescriptions to other planets (or exoplanets) is to stand on thin ice. Worryingly, there is an emerging sub-community of researchers switching over to exoplanet science, from the Earth sciences, who are bringing with them such Earth-centric approaches. 

To form large-scale galaxies, one needs prescriptions for star formation and how dying stars (supernovae) feed energy back into their environments. To simulate the climate, one needs prescriptions for turbulence and precipitation. Such prescriptions often employ a slew of free parameters that are either inadequately informed by data or involve poorly known physics. As the number of free parameters in a simulation increase, so does the diversity and variety of simulated results. In the most extreme limit, the simulation predicts everything---it is consistent with every outcome anticipated. A quote attributed to John von Neumann describes it best, ``With four parameters, I can fit an elephant and with five I can make him wiggle his trunk." Inattention to falsifiability has been chided by Wolfgang Pauli, who remarked, ``It is not only incorrect, it is not even wrong." \textit{(``Das ist nicht nur nicht richtig, es ist nicht einmal falsch.")} A simulation that cannot be falsified can hardly be considered science. 

Simulations as a third way of establishing scientific truth are here to stay. The challenge is for the astrophysical community to wield them as transparent, reproducible tools, thereby placing them on an equally credible footing with theory and experiment.

\vspace{0.15in}
\noindent
\textbf{Acknowledgment}
\vspace{0.05in}

The author is grateful to Scott Tremaine, George Lake and Justin Read for constructive feedback on an earlier version of the essay.

\label{lastpage}


\begin{thebibliography}{99}

\bibitem[Dyson(2004)]{dyson} Dyson, F., 2004., ``A meeting with Enrico Fermi", \textit{Nature}, 427, 297

\bibitem[Held(2005)]{held} Held, I.M., 2005, ``The gap between simulations and understanding in climate modeling", \textit{Bulletin of the American Meteorological Society}, 86, 1609

\bibitem[Ostriker(1997)]{ostriker} Ostriker, J.P., 1997, ``Obituary: Martin Schwarzschild (1912-97)", \textit{Nature}, 388, 430

\bibitem[Poincare(2001)]{poincare} Poincar\'{e}, H., 2001, ``The Value of Science: Essential Writings of Henri Poincare", edited by Stephen Jay Gould (New York: Modern Library)


\end{thebibliography}
\end{document}